\documentclass[conference]{IEEEtran}
\IEEEoverridecommandlockouts
\usepackage{cite}
\usepackage{amsmath,amssymb,amsfonts}
\usepackage{algorithmic}
\usepackage{graphicx}
\usepackage{textcomp}
\usepackage{comment}
\usepackage{xcolor}
\usepackage{booktabs} 

\usepackage{subcaption} 

\newcommand{\etal}{\textit{et al}., }
\def\BibTeX{{\rm B\kern-.05em{\sc i\kern-.025em b}\kern-.08em
    T\kern-.1667em\lower.7ex\hbox{E}\kern-.125emX}}

\begin{document}

\title{Security Risks Due to Data Persistence in Cloud FPGA Platforms
}

\author{\IEEEauthorblockN{Zhehang Zhang, Bharadwaj Madabhushi, Sandip Kundu, and Russell Tessier}
\IEEEauthorblockA{\textit{Department of Electrical and Computer Engineering} \\
University of Massachusetts, Amherst, MA, USA\\
\{\textit{zhehangzhang, bmadabhushi, kundu, tessier}\}@\textit{umass.edu}}
}
\maketitle

\begin{abstract}
The integration of Field Programmable Gate Arrays (FPGAs) into cloud computing systems has become commonplace. As the operating systems used to manage these systems evolve, special consideration must be given to DRAM devices accessible by FPGAs. These devices may hold sensitive data that can become inadvertently exposed to adversaries following user logout. Although addressed in some cloud FPGA environments, automatic DRAM clearing after process termination is not automatically included in popular FPGA runtime environments nor in most proposed cloud FPGA hypervisors. In this paper, we examine DRAM data persistence in AMD/Xilinx Alveo U280 nodes that are part of the Open Cloud Testbed (OCT). Our results indicate that DDR4 DRAM is not automatically cleared following user logout from an allocated node and subsequent node users can easily obtain recognizable data from the DRAM following node reallocation over 17 minutes later. This issue is particularly relevant for systems which support FPGA multi-tenancy.
\end{abstract}


\section{Introduction}\label{section:Introduction}

FPGA use in the cloud is substantial and likely to grow as the spectrum of cloud applications increases \cite{shahzad2021survey}. Cloud FPGAs generally are attached to substantial DRAM storage that is either directly attached to the devices or accessible via a system-level bus \cite{bobda2022future} (e.g., PCIe). Current cloud FPGA vendors allocate a single user to an FPGA at a time (single-tenant), although multi-tenant use in which multiple independent users share FPGA logic and attached memory at the same time \cite{mbongue2021deploying} has been proposed. In both scenarios, FPGAs and attached DRAM are used by numerous untrusted users. DRAM data should obviously stay confidential for individual users. However, unlike cloud CPUs that are often allocated as virtual machines managed by operating systems with sophisticated memory management such as Linux, cloud FPGA resources often are managed at a lower level. As we describe in this paper, simply clearing an FPGA logic configuration upon user deallocation does not immediately erase its data in the attached DRAM. These memory locations must be explicitly cleared when a user is deallocated memory space. Although explicit DRAM clearing is used by some existing cloud FPGA platforms (notably AWS EC2 F1 \cite{bobda2022future}), it is not universally deployed in all cloud platforms.

In this paper we comprehensively examine DRAM data persistence in AMD/Xilinx Alveo U280 platforms found in the publicly-available Open Cloud Testbed (OCT) \cite{zink2021open}. We show that even though a user's FPGA configuration information, including a DRAM controller, is removed from the device upon system logout, usable DRAM data persists for well over twenty minutes even if DRAM refresh is not performed. This data is accessible to subsequent users after the FPGA is reallocated to another user. The issue is even more significant in the case of multi-tenancy when DRAM use by one tenant may lead to refreshes of stale data from previous users that has not yet been erased. These issues are explored in the context of OCT and the Xilinx Runtime Library (XRT).

The remainder of the paper is structured as follows. In Section \ref{sec:Background}, we discuss how bulk memory is typically accessed by cloud FPGAs, with a focus on DRAM that is directly attached to the devices. We also discuss OCT and how runtime management software initiates DRAM access and manages FPGA resources. In Section \ref{sec:approach}, we describe several experiments which lead to the retention of data values written into DRAM following FPGA reconfiguration. We draw parallels between the cloud FPGA architecture in OCT and similar cloud FPGA architectures. In Section \ref{sec:results}, experimental results are presented
showing data persistence in OCT DRAM attached to FPGAs. We conclude the paper in Section \ref{sec:conclusion}.

\section{Background}\label{section:Background}
\label{sec:Background}
\subsection{DRAM data retention}
Cloud computing platforms contain a wide variety of computing components such as microprocessors (CPUs), graphics processing units (GPUs), and FPGAs. CPU-based systems, including the platforms which serve as computing node hosts in the cloud, typically use Linux- or Windows-based operating systems (OSs) to allocate DRAM and other memory attached to
the processor. Individual DRAM pages are explicitly erased prior to
reallocation ensuring that retained data from an earlier process cannot
be snooped by a subsequent process \cite{tannenbaum2009}. This protection is not necessarily present in
Linux-based systems-on-chip that include
both CPU and FPGA logic, as a recent memory-scraping attack showed \cite{Madabhushi2024date}.

GPUs can suffer from memory persistence issues \cite{lee14stealing, zhou2017pets}. The application programming interfaces
(APIs) associated with GPUs offer significant sharing but often do
not initialize newly-allocated memory. These weaknesses have been
enumerated and successful attacks have obtained portions of persistent web pages and other images \cite{zhou2017pets} following user logout.
Suggested remediations include user-initiated scrubbing of memory
following use. Similar issues affect ARM Mali GPUs, where new processes can access terminated process pages due to the reuse of freed pages \cite{gh}.

Several operating systems and hypervisors have been introduced that time multiplex one or more user circuits into reconfigurable regions in cloud FPGA devices. Each user circuit (often called a kernel) is generally allocated a logic region and portions of attached and bus-accessed DRAM. For example, OPTIMUS \cite{ma2020optimus} allocates FPGA logic and data pages on demand. Although internal FPGA state is cleared after use, no discussion of DRAM clearing is provided. In Byma et al. \cite{byma2014fpgas} address space is partitioned across tasks, effectively creating private pages and in Ruan et al. \cite{ruan2022fpl} allocation is controlled by an ARM processor. Finally, Korolija et al. \cite{Korolija20Abstractions} restrict inter-kernel access to DRAM using translation lookaside buffers (TLBs). None of these systems indicate DRAM clearing upon kernel clearing or DRAM reallocation.

\subsection{Experimental Platform}
\begin{figure}[t]
    \centering
    \includegraphics[width=\linewidth]{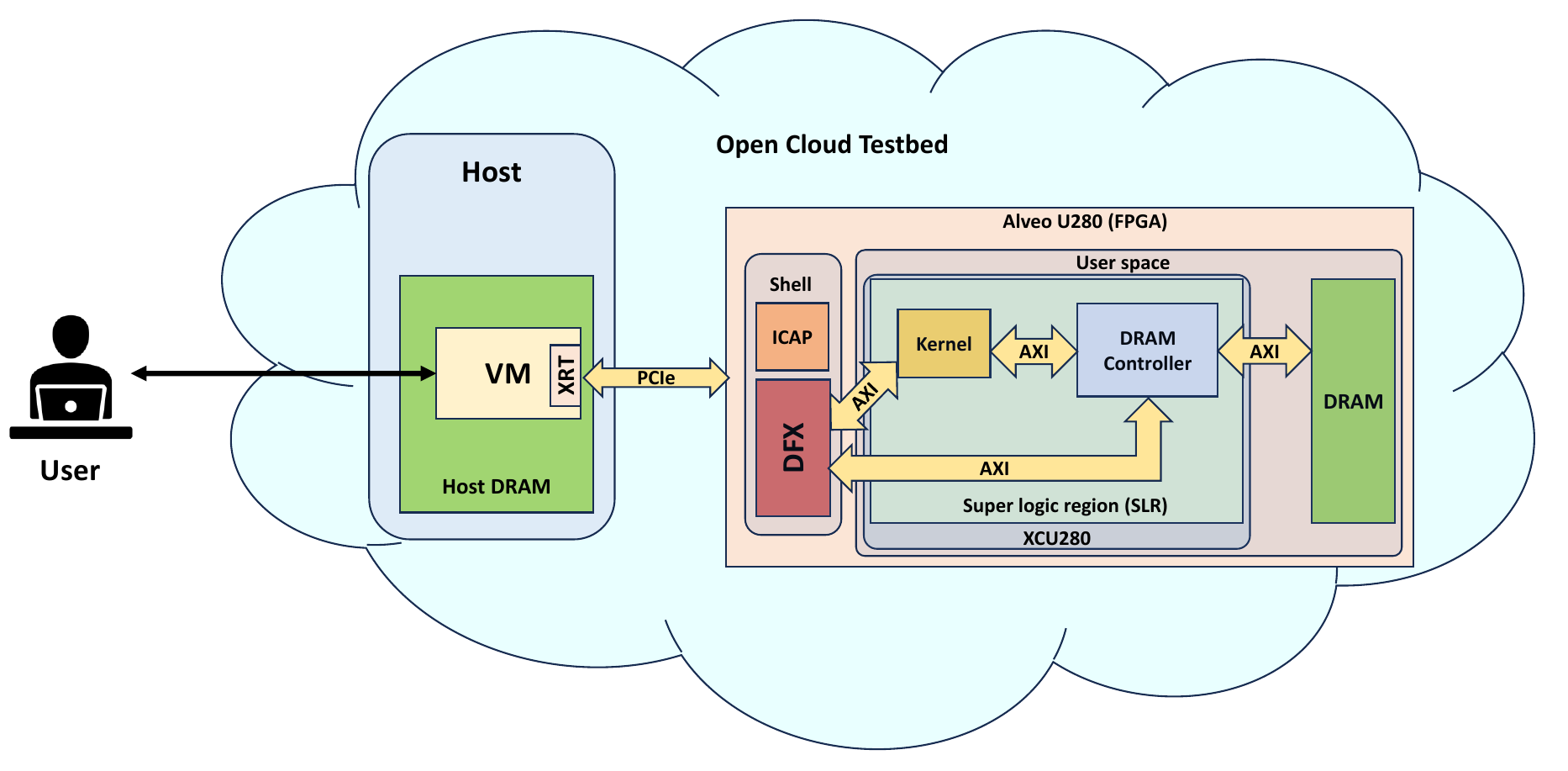}
    \caption{Overview of AMD Alveo U280 integration in OCT}
    \label{fig:FPGA_in_OCT}
\end{figure}

In this paper, we examine the disposition of FPGA-connected DRAM following user logout under typical use. Our experimentation is performed
using the publicly-available OCT, which includes CPUs, GPUs, and FPGA  accelerators. The OCT host processor executes a version of Linux and supports virtual machines. Only one user is allocated to a node at a time. The Alveo U280 data center accelerator cards \cite{u280} in OCT use PCIe
connections to the host processor and are also directly
connected to a network switch via two independent
100Gbps connections \cite{Handagala2021cnert}.  The
configuration of an FPGA node within the OCT and its internal logical connections are depicted in Figure
\ref{fig:FPGA_in_OCT}. It should be noted that similar Alveo U250 cards are integrated into some
Microsoft Azure nodes \cite{azure}. 

The host processor in OCT interfaces to the XCU280 UltraScale+ FPGA via APIs provided by the Xilinx Runtime Library. The XRT environment 
facilitates FPGA configuration and runtime data transfer between the
host and an FPGA in a variety of cloud and embedded computing systems. 
As shown in Figure \ref{fig:FPGA_in_OCT}, in OCT, the FPGA's PCIe
connection is created using a dynamic function exchange (DFX)
interface. This circuitry, which is loaded into the FPGA from flash,
accepts data from the PCIe bus to configure FPGA kernels (user
circuits) via ICAP and transfers data to FPGA memory (attached DRAM, embedded
high bandwidth memory (HBM) and programmable logic RAM (PLRAM)). 
FPGA DRAM is accessed via a DRAM controller. The controller is attached to
an AXI bus connected to the DFX interface and kernel circuitry. In general, XRT
and the similar Open Programmable Acceleration Engine (OPEA) \cite{opae} for Intel
FPGA fabrics do not natively clear DRAM attached to FPGAs following
process termination, leaving it to the user to clear memory resources. 

\subsection{DRAM Decay}
The DRAM attached to the XCU280 FPGA in Figure \ref{fig:FPGA_in_OCT} is controlled by a DRAM controller fashioned from FPGA logic that is only present when the associated controller circuit is instantiated in the device. In addition to write and read requests, this controller generates signals to perform auto refresh for the two 16GB Micron MTA18ASF2G72PZ DDR4 SDRAM RDIMMs \cite{dram} in the U280. If the FPGA is reset (configuration cleared) or a new FPGA configuration is programmed into the device that does not contain a DRAM controller, refresh signals will cease, and stored DRAM values will decay. For older DDR1 and DDR2 devices \cite{gruhn2013, halderman2009lest}, this decay is generally quite rapid (on the order of seconds). Experiments with more recent DRAM technology measured extended decay that ranges into minutes \cite{rahmati2014refreshing}. However, these experiments do not consider cloud FPGA interfacing and more modern DDR4 technology. 

It should be noted that ``charged" DRAM cells may be read as either a logic 1 or a logic 0 depending on DRAM device architecture.
Our research shows that half the initially charged cells in the DDR4 chips in the U280 will decay from a logic 1 to a logic 0 and half will decay from a logic 0 to a logic 1, similar to observations in \cite{xiong2019date}. This characterization can be performed by an attacker prior to performing an attack by programming the DRAM with all zeros or ones and observing decay patterns after re-logging into the same node.

\section{Experimental Approach}\label{section:experimentalsetup}
\label{sec:approach}

The bitstreams used in our experimentation were generated using AMD Vitis version 2023.1 which produces bitstreams for the Alveo U280
hardware. The OCT host computer includes an Intel Xeon Gold 6226R CPU  operating at 2.90GHz with 187GB of memory.

To determine DDR4 DRAM data retention in the byte-addressed Alveo U280 following the loss of refresh capabilities, a series of experiments were performed to write data into the DRAM, remove DRAM refresh in auto refresh mode, and resample the DRAM data following specific time periods. In all experiments (except where noted), each DRAM cell was initially charged. As noted in the previous section, half the cells stored a logic 1 which decayed to 0 and half stored a logic 0 which decayed to a 1. For example, if byte address location 0x00000000 decays to 0x00, we write 0xFF, and vice versa for addresses with the opposite polarity. The polarity of each cell was determined prior to experimentation. Our data retention results were generated using two types of experiments:

{\bf Experiment 1 - Session termination:} In this experiment, a user logs into an OCT node and uses an FPGA kernel and DRAM controller to write 4GB of data to DRAM bank DDR\_0. After these writes are complete, the user logs out of the node, causing an immediate FPGA warm reset (and corresponding configuration and DRAM controller clear). Once the node is reset and free, any user (including the one that just logged out) can attempt to reclaim it. After re-login success, the FPGA is configured with a new user kernel and a DRAM controller and the contents of DDR\_0 are retrieved by the host via the AXI and PCIe busses. As noted later in this section, the average time from node release via logout to re-login is about 17.25 minutes. This experiment mimics a realistic attack since the second user could be anyone, including an attacker. Effectively, this value measures the amount of time between when one user logs out and the next user can log in.

{\bf Experiment 2 - DRAM controller removal and subsequent reinsertion:} Although Experiment 1 is useful for examining data persistence for a attack involving multiple independent node logins, it does not allow for examination of data persistence for time spans of less than 18 minutes. Thus, we constructed a second experiment that controls the length of DRAM decay. Although this experiment does not mimic an actual attack, it provides information on data persistence during the time period between logout and re-login. In this experiment, the user logs into an OCT node and uses an FPGA kernel and DRAM controller to write 4GB of data to DRAM bank DDR\_0. After these writes are complete, the FPGA is reconfigured to remove the DRAM controller, effectively ending DRAM refreshes. After a user-determined period of time, the FPGA is reconfigured with a new user kernel and a DRAM controller and the contents of DDR\_0 are retrieved by the host via the AXI and PCIe busses. The original user does not log out of the node during this experiment.
We performed a series of trials using this approach in which the DRAM values were sampled with an increasing wait time of one minute per trial (e.g., one minute, two minutes, three minutes, etc., up to 18 minutes). 

For both experiments, it is straightforward for an attacker to determine which DRAM cells in a node decay. By taking the XOR of the reference data and the read data after the attacker logs back into the same node it is possible to find the faulty bit locations. In a subsequent attack, an attacker can log into the same node immediately after a victim uses it and access the DRAM bits that have been previously identified as unlikely to decay.

It should be noted that DRAM readback from bank DDR\_0 cannot be performed directly following FPGA reconfiguration in Experiments 1 and 2. XRT requires that data be copied from bank DDR\_0 to bank DDR\_1 first before being read by the host via the AXI and PCIe busses. However, this step does not impact the values read from DDR\_0 since DDR\_1 simply serves as an intermediate buffer. The 4GB DRAM readback required 336 ms on average in our experimentation.

Our experimentation was performed on four separate OCT FPGA-based nodes. Each node produced similar results for minimum logout to re-login time for Experiment 1. For example, across four trials, this time gap ranged from 16 minutes, 58 seconds to 17 minutes, 53 seconds, with an average of 17 minutes, 24 seconds. Therefore, an attacker who has been monitoring a victim and waits for the victim to terminate their connection to an Alveo U280 in OCT can reconnect to the same U280 in less than 18 minutes and inspect any leftover DRAM values. This time gap includes time for the OCT management software to clear the victim's virtual machine, create a new virtual machine for the attacker, boot the new virtual machine, install the XRT tools, and establish a connection for the attacker. 

\section{Results}\label{section:results}
\label{sec:results}

\subsection{Data Decay Analysis}

\begin{figure}[t]
    \centering
    \includegraphics[width=\linewidth]{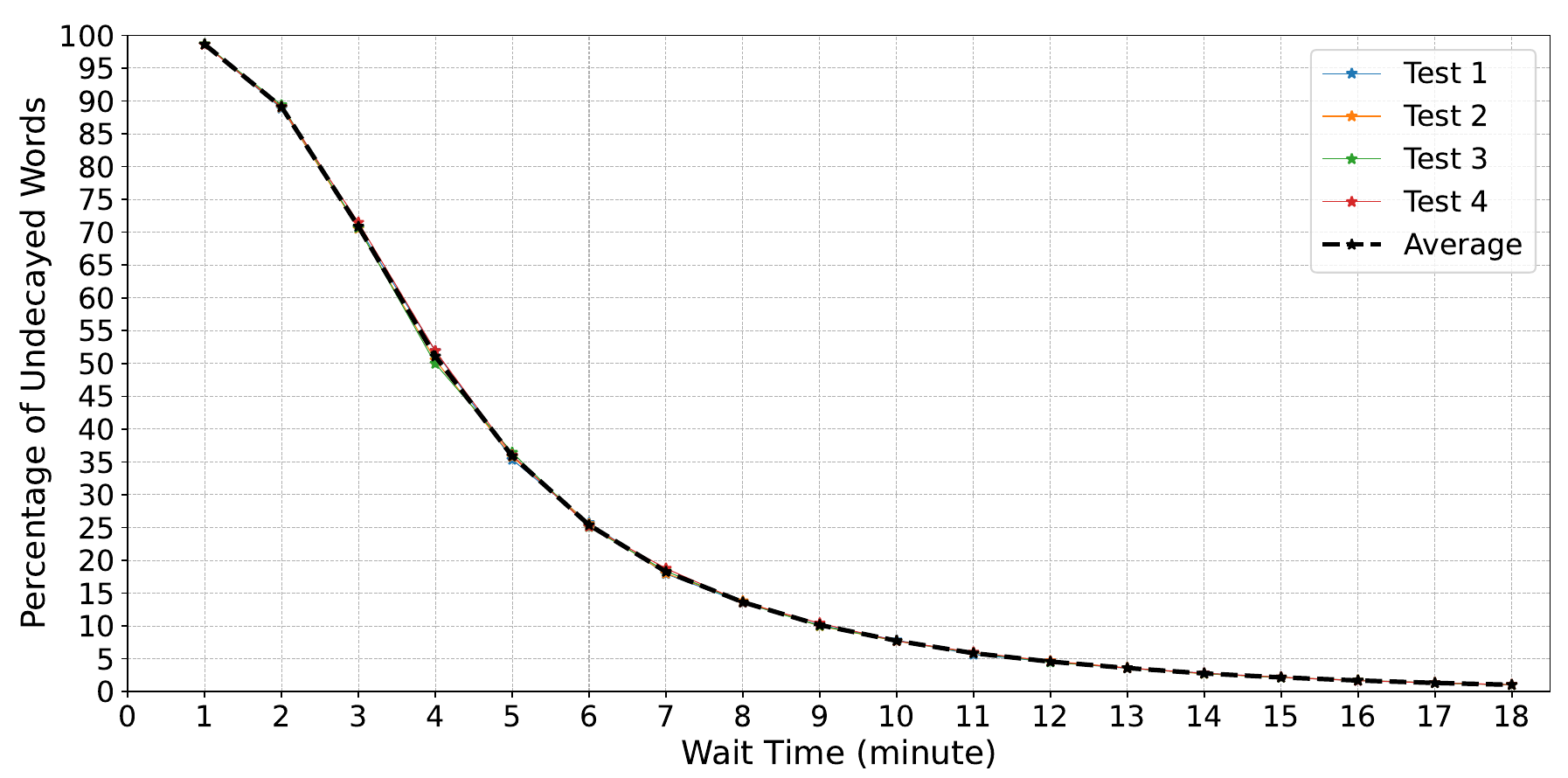}
    \caption{Decay rate of 32-bit words in OCT node PC151. Decay indicates a change in value of any of the 32 bits of a word. Total memory size is 4 GB.}
    \label{fig:undecayrandomdata}
\end{figure}

\begin{figure}[t]
    \centering
    \includegraphics[width=\linewidth]{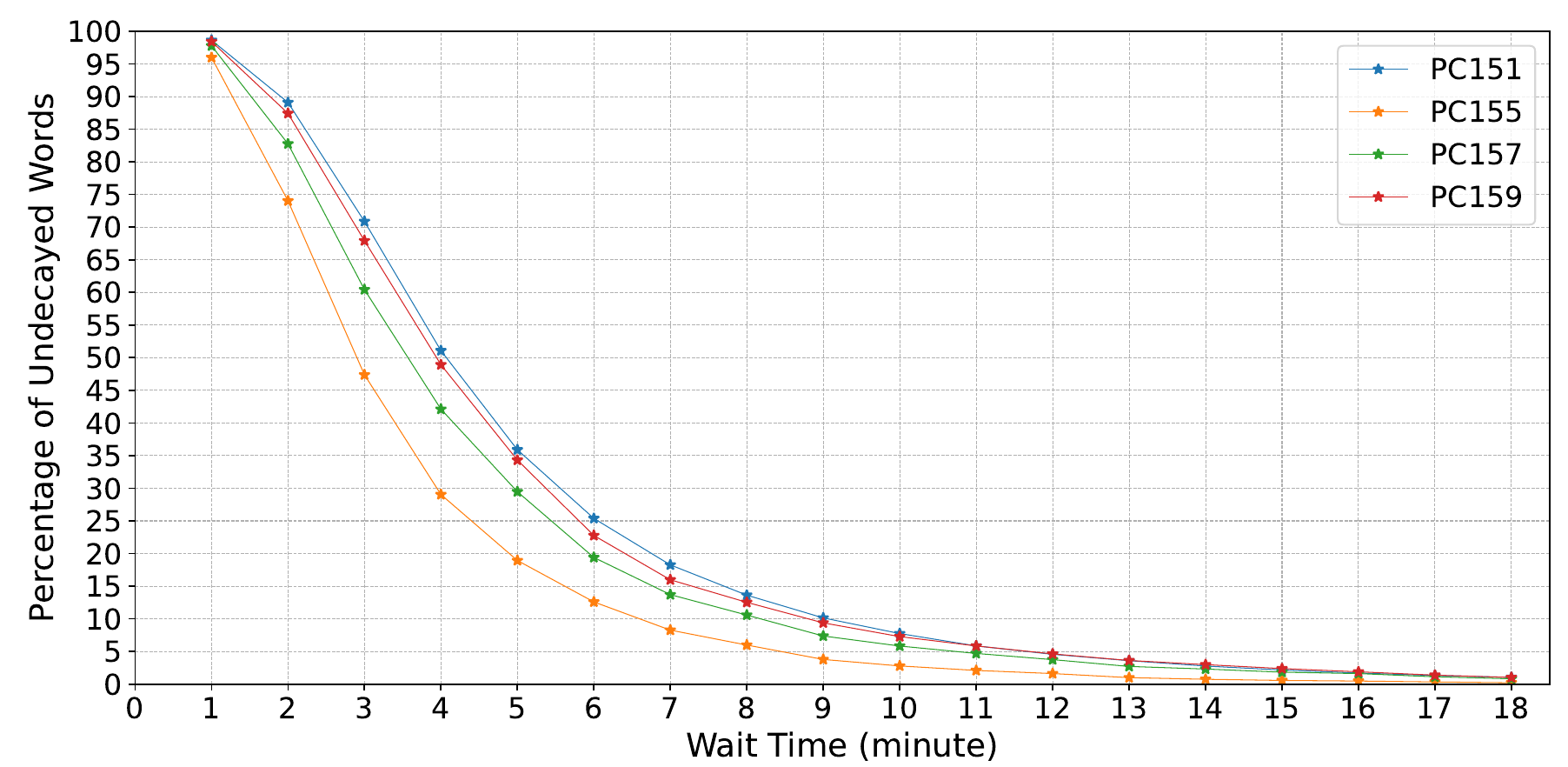}
    \caption{Decay rate of 32-bit words for four OCT nodes averaged over four trials. Decay indicates a change in value of any of the 32 bits of a word. Total memory size is 4 GB.}
    \label{fig:undecayrandomdata_all}
\end{figure}

In this section we examine the decay of DRAM data at one minute intervals from one minute to 18 minutes (re-login delay). Experiment 2 was used to generate all data except the last data point (Experiment 1) in our plots. Our analysis considers the decay of words (32 bits) and individual bits in a 4GB portion of bank DDR\_0. We consider a 32-bit word to be decayed if any of the bits have changed from their original values. Figure \ref{fig:undecayrandomdata} shows decay percentages of values over four sets of trials for OCT Alveo U280 nodes PC151, indicating consistency across trials. The average decay rates across the four OCT nodes mentioned in the previous section are shown in Figure \ref{fig:undecayrandomdata_all}.

\begin{figure}[t]
    \centering
    \includegraphics[width=1\linewidth]{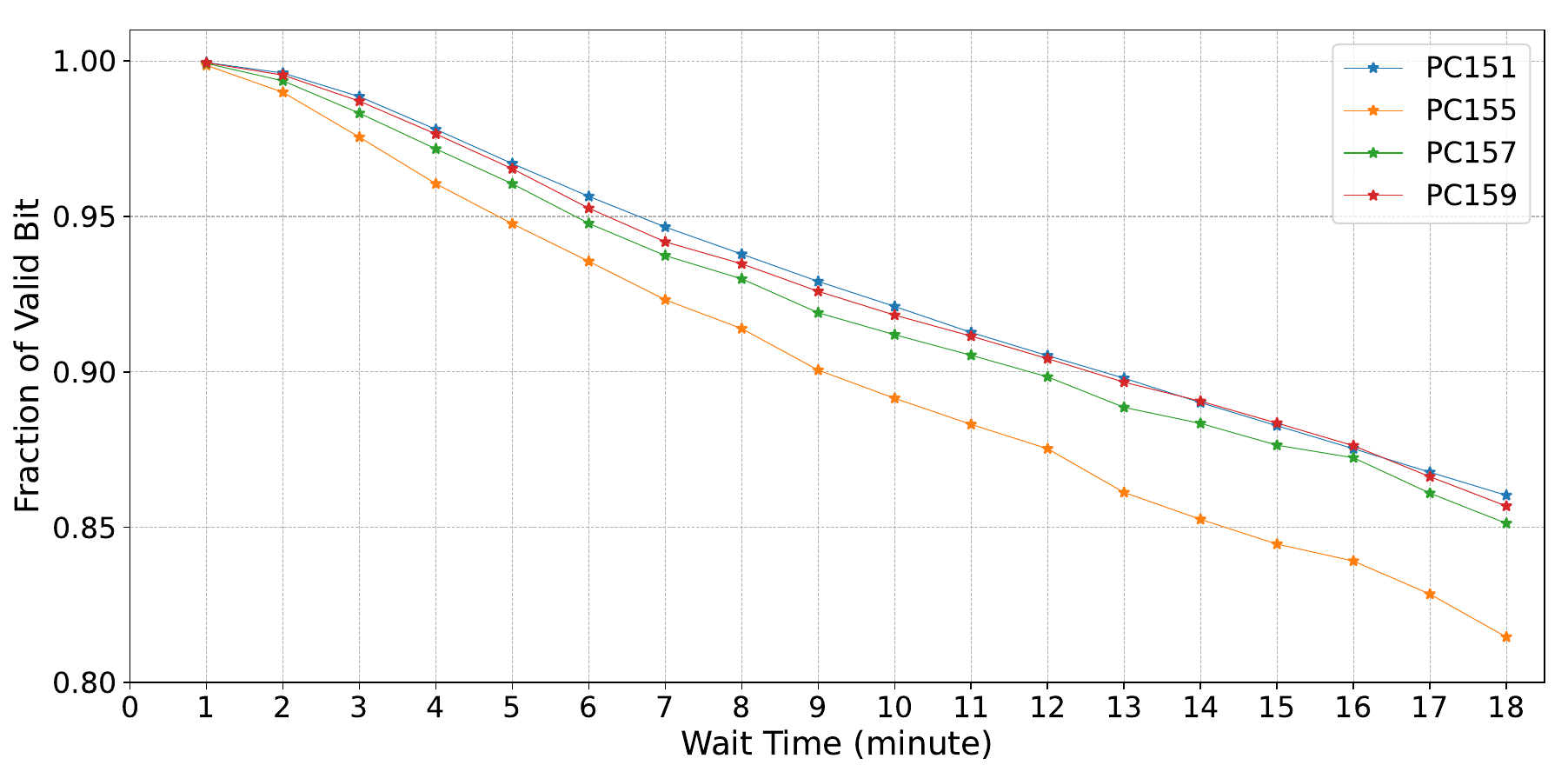}
    \caption{Fraction of valid individual bits in four OCT nodes over time. Total memory size is 4 GB.}
    \label{fig:bitwidthundecay}
\end{figure}

The decay percentages indicate that a substantial fraction of the words in the DRAM (between 30 and 50\%) remain unchanged four minutes after refreshes are terminated. However, by the 18 minute mark (re-login delay) less than 1\% of 32-bit values are unaffected. Despite this small percentage, more than one million byte address locations are unaffected. The addresses of the unaffected data remained consistent across trials. For example, on PC151, 95.45\% of the addresses of unaffected data remained consistent from trial to trail while the number was 95.79\% for PC157.
Although there is a significant decay in word data, the integrity of single-bit data remains considerably better. After 18 minutes using Experiment 1, as demonstrated in Figure \ref{fig:bitwidthundecay}, 86\% of bits remain valid (undecayed). 

Table \ref{tab:percentageof0f} provides an additional data point regarding valid data at the 18-minute mark. The table illustrates the percentage of nibbles that remain undecayed per 32-bit word. The column sums to 100\%. Notably, the table demonstrates a significant preservation of original data states, with approximately 55\% of words retaining at least half of their nibbles.


\begin{table}[t]
\centering
\begin{tabular}{cr}

\hline
\# of Undecayed Nibbles & Percentage\\
\hline
8 & 1.5\%\\
7 & 7.9\%\\
6 & 18.8\%\\
5 & 26.3\%\\
4 & 23.7\%\\
3 & 14.3\%\\
2 & 5.7\%\\
1 & 1.5\%\\
0 & 0.1\%\\
\hline
\end{tabular}
\caption{Percentage of undecayed nibbles in each 32-bit value at the 18-minute decay point}
\label{tab:percentageof0f}
\end{table}

\begin{figure}[t]
  \centering
  \begin{subfigure}[b]{0.48\linewidth}
    \includegraphics[width=\linewidth]{input_1-1.pdf}
    \caption{Initial}
    \label{fig:sub1}
  \end{subfigure}
  \hfill 
  \begin{subfigure}[b]{0.48\linewidth}
    \includegraphics[width=\linewidth]{output_1.pdf}
    \caption{Readback after 18 minutes}
    \label{fig:sub2}
  \end{subfigure}
  \caption{Contrast between the initial image and the image read back via Experiment 1 after 18 minutes}
  \label{fig:Contrast}
\end{figure}

In a final experiment, we performed Experiment 1 using a 3,344$\times$5,016 RGB image. Each 32-bit pixel includes eight bits each of red, green, and blue and eight bits of alpha (opaqueness). The original image \cite{reding_2024} is shown in Figure \ref{fig:Contrast}a. The decayed version, retrieved after 18 minutes on PC151 is shown in Figure \ref{fig:Contrast}b. Except for discoloration and some distortion, the image is still recognizable. It should be noted that after 18 minutes, 85\% of bits remain unchanged in DRAM, according to Figure \ref{fig:bitwidthundecay}. Bit retention values varied across eight bit channels (red: 83\%, green: 80\%, blue: 82\%, alpha: 96\%) leading to a purplish hue in the recovered picture. Since alpha values are typically 0xFF rather than random values, their retention values were higher.

\subsection{Observations}
\begin{itemize}
\item The DRAM controller in the AMD Alveo series of accelerators is implemented as a soft IP. When the FPGA is reset, the DRAM controller is deleted and the DRAM no longer refreshes.
\item Consequently, data in the FPGA's local DRAM begins to decay. The decay rate is different for different bits, with some bits retaining data beyond 18 minutes, allowing attackers to access these values.
\item Additionally, we observe that address locations exhibit consistent decay rates over time. This implies that specific address locations within an FPGA's DRAM decay at the same rate. For instance, shortly after a warm reset is applied, certain address locations decay more rapidly than others, while some decay at a slower pace, and so forth.
\item In a multi-tenant environment, the DRAM controller persists in an active state even after a process termination to service other ongoing processes. This issue allows a full read of the previous user's data from the FPGA's DRAM by a subsequent user allocated the previous user's address space.
\item These findings extend to other FPGAs that use a soft-core DRAM controller.
\end{itemize}

\subsection{Other Comments}

DRAMs, including the chips used on the Alveo U280, generally support a self-refresh mode which allows the memory chips to refresh values even in the absence of an external refresh signal. This mode could be abused to retain fully valid data indefinitely if power to the DRAM is maintained after FPGA reconfiguration. However, the U280 appears to assert a reset signal to DRAM if the FPGA configuration is cleared, which draws the DRAM out of self refresh mode. This prevents the use of self refresh data retention and requires the use of auto refresh.
\section{Conclusion}\label{section:conclusion}
\label{sec:conclusion}


This work investigates security vulnerabilities arising from FPGA DRAM usage in cloud environments. Experiments on the Open Cloud Testbed (OCT) platform, incorporating Alveo U280 boards, were conducted to assess DRAM data persistence after user access. The results underscore the critical need for DRAM data erasure following user sessions in cloud FPGAs, which serves as an effective countermeasure to our attack. Even with DRAM controller removal during a warm reset, residual data persists in various DRAM locations that a new user can access, despite an overhead of nearly 18 minutes to reconfigure a FPGA virtual machine on the OCT platform. This vulnerability is particularly concerning in multi-tenant environments, where auto-refresh can continuously maintain data integrity, potentially leading to indefinite data retention. These observations highlight the necessity for robust data shredding mechanisms to ensure data privacy and security in cloud FPGA deployments.
\section{Acknowledgments}

This research was supported in part by National Science Foundation grant CNS-2247059. 
We are grateful to the Open Cloud Testbed for complimentary access, which was instrumental in producing the results for this paper.
\bibliographystyle{IEEEtran}
\bibliography{crypto}

\end{document}